\documentclass{mynat}
\usepackage{graphicx}

\title{A \boldmath{$15.65\,M_{\odot}$}
black hole in an eclipsing binary in the nearby spiral galaxy Messier 33}

\author{Jerome A. Orosz$^{1}$, 
Jeffrey E. McClintock$^2$, 
Ramesh Narayan$^2$,
Charles D. Bailyn$^3$,
Joel D. Hartman$^2$,
Lucas Macri$^4$,
Jiefeng Liu$^2$, 
Wolfgang Pietsch$^5$, 
Ronald A. Remillard$^6$,
Avi Shporer$^7$ \&
Tsevi Mazeh$^7$
}

\def\lesssim{\mathrel{\hbox{\rlap{\hbox{\lower4pt\hbox{$\sim$}}}\hbox{$<$}}}}
\def\gtrsim{\mathrel{\hbox{\rlap{\hbox{\lower4pt\hbox{$\sim$}}}\hbox{$>$}}}}
\def\ggg{\mathrel{\hbox{\rlap{\hbox{\lower4pt\hbox{$\sim$}}}\hbox{$>$}}}}

\begin{document}

\maketitle

\begin{affiliations}
\item Department of Astronomy, San Diego State University,
5500 Campanile Drive, San Diego, CA 92182-1221, USA.
\item Harvard-Smithsonian Center for Astrophysics, 60 Garden Street,
Cambridge, MA 02138, USA.
\item Department of Astronomy, Yale University,
PO Box 208101, New Haven, CT 06520-8101, USA.
\item National Optical Astronomy Observatory, 950 North
Cherry Avenue, Tuscon, AZ 85719, USA.
\item Max-Planck-Institut f\"ur extraterrestrische Physik,
Giessenbachstra\ss e,  D-85741 Garching, Germany.  
\item MIT Kavli Institute for Astrophysics and Space Research, 
77 Massachusetts Avenue, 37-287, 
Cambridge, MA 02139, USA.
\item Wise Observatory, Tel Aviv University, Tel Aviv 69978, Israel.  
\end{affiliations}

\begin{abstract}
Stellar-mass black holes are discovered in X-ray emitting binary
systems, where their mass can be determined from the dynamics of their
companion stars\cite{rem06,cha06,oro03}.  Models of stellar evolution
have difficulty producing black holes in close binaries with masses
\boldmath{$>10\,M_{\odot}$} (ref.\ 4), which is consistent with the
fact that the most massive stellar black holes known so
far\cite{cha06,oro03} all have masses within \boldmath{$1\sigma$} of
\boldmath{$10\, M_{\odot}$}. Here we report a mass of \boldmath{$15.65
\pm 1.45\,M_{\odot}$} for the black hole in the recently discovered
system M33 X-7, which is located in the nearby galaxy Messier 33 (M33)
and is the only known black hole that is in an eclipsing binary\cite{pie06}.  
In order to produce such a massive black hole,
the progenitor star must have retained much of its outer envelope
until after helium fusion in the core was completed\cite{bro01}.  On
the other hand, in order for the black hole to be in its present 3.45 day orbit
about its \boldmath{$70.0 \pm 6.9 M_{\odot}$} companion, there
must have been a ``common envelope'' phase of evolution in which a
significant amount of mass was lost from the system\cite{tau06}.  We
find the common envelope phase could not have occured in M33 X-7
unless the amount of mass lost from the progenitor  during
its evolution was an order of magnitude less than what is usually
assumed in evolutionary models of massive
stars\cite{sch92,mey94,vaz07}.
\end{abstract}

Optical imaging and spectroscopic observations of M33 X-7 were
obtained in service mode with the 8.2m Gemini North Telescope between
2006 August 18 and November 16.  The mean optical spectrum is shown in
Figure 1.  The radial velocities derived from the 22 usable spectra
show a nearly sinusoidal variation when phased on the orbital period
of 3.453014 days determined from the X-ray eclipses\cite{pie06} (Fig.\
2b).

Time series photometry was derived from the Gemini images in the Sloan
g$^{\prime}$ and r$^{\prime}$ filters (Fig.\ 3).  Additional
photometric data were obtained in the $B$, $V$, and $I$ filters using
the 3.5m WIYN telescope during 2006 August 18-21 and 2006 September
15-16 (see Supplementary Information).  The phased light curves show
the characteristic ellipsoidal variations of a tidally distorted star
(Fig.\ 3), which have been reported previously for this
source\cite{pie04,shp07}.

The temperature of the companion star was determined by comparing its
averaged spectrum (Fig.\ 1) to a collection of synthetic spectra
derived from the OSTAR2002 grid\cite{lan03}.  As is usually the case,
there is a strong correlation in the library of models between
effective temperature, $T_{\rm eff}$, and surface gravity, $\log g$,
and thus various combinations of these parameters result in very
similar spectra.  Fortunately, the dynamical information strongly
constrains the allowed value of $\log g$ to be between 3.65 and 3.75
at the $3\sigma$ level.  A good match to the observed spectrum is
provided by the model with $T_{\rm eff}=35000$~K, $\log g=3.75$, a
metalicity of 10\% of the solar value (representative of star clusters
in M 33 at this galactocentric distance\cite{ma01}), and our measured
value of the projected rotational velocity of $V_{\rm rot}\sin i=250
\pm 7$ km s$^{-1}$ (where $i$ is the orbital inclination angle).  The
formal error on the temperature is $\pm 200$~K for $\log g=3.75$.
However, given the possibility of a gravity slightly different than
$\log g=3.75$ (the grid spacing of the models is 0.25 dex) and the
correlation between $T_{\rm eff}$ and $\log g$, we adopt a temperature
in the range $34000\le T_{\rm eff}\le 36000$~K, which corresponds to a
spectral type of O7III to O8III\cite{hea07}.

To compute the radius of the O-star, we adopt a distance modulus of
$24.62\pm 0.05$ mag ($d=840\pm 20$~kpc) to M33 (see the Supplementary
Information for details).  Using $V = 18.9\pm 0.05$, $A_V = 0.53\pm
0.06$ mag (ref.\ 10), and the bolometric corrections derived from the
OSTAR2002 grid, we find a radius of $R_2=19.6\pm 0.9 \,R_{\odot}$ and
a luminosity of $\log (L/L_{\odot})=5.72\pm 0.07$.

The duration of the X-ray eclipse $\Theta$ (measured in degrees of
orbital phase) strongly constrains the available parameter space.
Because the size of the compact X-ray source ($\lesssim 1000$ km) is
vastly smaller than the secondary star, one might expect the X-ray
eclipse profile to be a ``square well'' with a flat bottom and very
abrupt periods of ingress and egress.  However, in M33 X-7 and other
X-ray binaries with massive companions, the observed eclipse profile
deviates from this idealised picture.  The transitions into and out of
eclipse are more gradual because of absorption of the X-rays in the
stellar wind that thickens near the O-star.  The non-zero X-ray
intensity in full eclipse results from X-rays that are scattered by
this wind around the O-star into our line-of-sight.  The erratic X-ray
variability prior to eclipse (Fig.\ 2a) is presumably caused by
absorption in the gas that is streaming from the O-star to the black
hole.  The period of egress is free of such effects, and we focus our
attention there.  We identify the eclipse width of $\Theta=53\pm
2.2^{\circ}$ (ref.\ 5) as the onset of the steep egress feature (solid
line in Fig.\ 2a).  In the Supplementary Information, we show that the
eclipse width of $\Theta=53^{\circ}$ is consistent with absorption in
the stellar wind, whereas the true eclipse by the stellar photosphere
corresponds to $\Theta=46\pm 1^{\circ}$
(dashed-dotted line in Fig.\ 2a).

We used a light curve synthesis code\cite{oro00} to find the optimal
model of the binary system.  
Fig.\ 3 shows the synthetic light curves for the best-fitting model,
which is schematically illustrated in Fig.\ 4.  The best-fitting model
parameters and derived astrophysical parameters are summarised in
Table 2.  The mass of the compact object is $M=15.65\pm
1.45\,M_{\odot}$, and the mass of the O-star is $M_2=70.0\pm
6.9\,M_{\odot}$, which puts it among the most massive stars whose
masses are well-determined\cite{gei03}.  The effective radii of the
Roche lobes are $21.8\,R_{\odot}$ and $10.8\,R_{\odot}$ for the O-star
and black hole, respectively.  From evolutionary models of {\em
single} stars\cite{sch92,mey94,vaz07} the age of the O-star is
estimated to be between about 2 and 3 million years.  We also note in
passing that the O-star is roughly a factor of three less luminous
than expected from the evolutionary models.

With $M = 15.65 \pm 1.45 M_{\odot}$, M33 X-7 is the most massive
stellar black hole known (see Table 1).  The mass of V404 Cyg is $12
\pm 2 M_{\odot}$ and the masses of the 18 other black holes, save one,
are $\lesssim 10 M_{\odot}$, or they are quite imprecise.  The one
contender is GRS 1915+105 with a mass of $14.0 \pm 4.4 \,M_{\odot}$
(refs.\ 17, 18).  However, the 30\% precision of the measurement is
poor.  Furthermore, there are reasons for questioning the reliability
of this impressive and pioneering result on a difficult system.  For
example, the spectroscopic orbital period\cite{gre01} is 9\% longer
than the recently-determined and precise photometric
period\cite{nei07}.  Furthermore, because of the large X-ray
luminosity, the late-type secondary star contributes ``only a few per
cent of the K-band brightness\cite{gre01};'' hence, the radial
velocity curve may be significantly distorted\cite{rey97}.  By
comparison, the mass estimate and eclipse ephemeris for M33 X-7 are
exceptionally precise, and X-ray heating is a minor effect.

M33 X-7 is a key system in the study of high mass stars, high mass
X-ray binaries, and high mass black holes.  A $\approx 16\,M_{\odot}$
black hole paired with a $\approx 70\,M_{\odot}$ secondary with a
separation of only $\approx 42 \,R_{\odot}$ is very difficult to
explain using stellar evolutionary models.  Since the radius of the black
hole progenitor would have been much larger than the current orbital
separation\cite{sch92,mey94,vaz07}, the two stars must have been
brought closer together via some kind of ``common envelope'' phase
which results in a significant amount of mass lost from the
progenitor, and very little mass gained by the secondary\cite{tau06}.
On the other hand, in order for the core mass to remain large enough
to produce a $\approx 16\,M_{\odot}$ black hole, the outer envelope of
the progenitor needs to be intact until core He burning is
completed\cite{bro01}.  Hence we require that the common envelope
phase begins only after core He burning in the progenitor is complete
(case C mass transfer\cite{tau06}).  There are two requirements for a
common envelope phase to start during case C mass transfer.  First,
the mass donor needs to be at least 1.2 times more massive than the
secondary at the start of mass transfer\cite{pod03}.  Second, the
radius of the mass donor at the end of core He burning needs to be
larger than its radius at the end of core H burning.  If the second
condition is not met, the CE phase begins {\em before} core He burning
is complete, and the stripped core loses much of its mass via strong
winds in its subsequent evolution and thus cannot make a massive black
hole\cite{bro01}.

Assuming no large change in the present-day mass loss rate of
$2.6\times 10^{-6}\,M_{\odot}\,{\rm yr}^{-1}$ (see Supplementary
Information), the secondary star has lost between about
$5.2\,M_{\odot}$ and $7.8\,M_{\odot}$, thereby putting its initial
mass near $\approx 80\,M_{\odot}$.  For a common envelope phase to
occur, the progenitor star should have been more massive than $\approx
80\times 1.2=96\,M_{\odot}$, which is problematic.  According to
evolutionary models\cite{sch92,mey94,vaz07}, massive stars lose much
of their initial mass via winds, and the mass loss rate generally
increases with increasing initial mass.  
For example, even in the extreme case of an initial mass of 
$120\, M_{\odot}$ and
a metallicity of 20\% solar, the mass after H burning is 
$\approx 52.9\,M_{\odot}$
and after He burning is $\approx 17.2\,M_{\odot}$ (ref. 9).
Furthermore, owing to the large amount of mass loss, the radius of the
star after core He burning is {\em smaller} than the radius after core
H burning.  For these reasons, a common envelope phase during case
C mass transfer seems very unlikely.  It would appear that the
progenitor star of M33 X-7 lost roughly an order of magnitude less
mass before the common envelope phase ensued than is predicted by the
evolutionary models.
Finally we note that there is an additional complication: even if a common
envelope is formed, the most likely outcome would be a
merger since the envelopes of massive stars are tightly
bound\cite{pod03}. However, the most massive star considered in ref.\
21 was $50\,M_{\odot}$, so the detailed computations should be
extended to the higher masses relevant for M33 X-7.

The determination of an accurate mass for M33 X-7 -- located at a
distance of more than 16 times that of any other confirmed stellar
black hole -- marks a major advance in our capability to study black
holes in Local Group galaxies beyond the Milky Way.

\begin{addendum}
 \item We thank J. Walsh for his help with the SPECRES software,
I. Hubeny for the use of his model atmosphere codes, and T. Matheson
for support with the Gemini Observations.
CDB acknowledges support from the
US National Science Foundation.
 \item[Competing Interests] The authors declare that they have no
competing financial interests.
 \item[Correspondence] Correspondence and requests for materials
should be addressed to J.A.O. \\ (email: orosz@sciences.sdsu.edu).
\end{addendum}

\spacing{1}

\begin{table}
\begin{tabular}{cccc}
\hline
X-ray source & Optical/IR counterpart & Black hole mass ($M_{\odot}$) 
  & Secondary star mass ($M_{\odot}$) \\
\hline
\hline
GRS 1915+105 & V1487 Aql & $14.0 \pm 4.4$ & $0.81\pm 0.53$ \\
GS 2023+338 & V404 Cyg & $12\pm 2$ & $0.6$ \\
A0620-00   & V616 Mon  & $10\pm 5$ & $0.6$ \\
GS 2000+25  & QZ Vul   & $10\pm 4$ & $0.5$ \\
XTE J1550-564 &  V381 Nor & $9.6 \pm 1.2 $  & ... \\
4U 1543-47 & IL Lup & $9.4 \pm 1.1$ & $2.5$ \\
Cyg X-1 & HDE 226868 & $>4.8$ & $>11.7$ \\
LMC X-1 & ...        & $8-20$ & ... \\
\hline
\end{tabular}
\caption{Recent dynamical 
measurements of massive stellar black holes.  Masses
for GRS 1915+105 are from ref.\ 18, and all others are taken 
from the compilation in ref.\ 2 and citations therein. 
The uncertainties correspond to one standard deviation.
There are three key ``observables'' that can be used to
determine the mass $M$ of the compact object in an X-ray binary.  
(1) The 
radial velocity semiamplitude
of the secondary star $K_2$, along with
the orbital period $P$ and eccentricity $e$, determines the mass function:
$
f(M)={PK_2^3}(1-e^2)^{3/2}/(2\pi G)= {M^3\sin^3i/(M+M_2)^2},
$
where
$M_2$ is the mass of the secondary star,
$i$ is the orbital
inclination angle and $G$ is the gravitational constant.  In order to
solve for $M$, we must determine $M_2$ (or $M_2/M$) and $i$, for which
we use (2) the rotational velocity of the secondary star $V_{\rm
rot}\sin i$, and (3) the amplitude of the ellipsoidal light curve.
The  two preceding observables
depend on $i$, $M_2/M$ and the Roche-lobe filling
$f_2$, which is the radial fraction of the secondary's Roche
equipotential lobe along the line of centers that is occupied by the star.
}
\end{table}

%
%

\begin{table}
\begin{tabular}{rrrrr}
\hline
parameter & value & \quad~\quad\quad~\quad\quad~\quad &  parameter & value \\
\hline
\hline
$\Theta$ (deg)     & $46\pm 1$  & & $M_2$ ($M_{\odot}$) & $70.0\pm 6.9$ \\
$T_{\rm eff}$ (K)  & $34000-36000$  & & $r_d$    & $0.45\pm 0.03$ \\
$V_{\rm rot}\sin i$ (km s$^{-1}$) & $250\pm 7$ & & $e$      & $0.0185\pm        0.0077$ \\
$R_2$ ($R_{\odot}$) & $19.6\pm 0.9$ & & $\omega$ (deg) & $140\pm 27$ \\
$\log L_2 $ ($L_{\odot}$) & $5.72\pm 0.07$ & & $\Omega$  & $0.903\pm 0.037$  \\
$\Delta\phi$   & $0.0045\pm 0.0014$   & & $f_2$       &  $0.777\pm 0.017$ \\
$i$ (deg) & $74.6\pm 1.0$ & & $a$ ($R_{\odot}$) & $42.4\pm 1.5$ \\
$K_2$ (km s$^{-1}$) & $108.9\pm 5.7$ & & $M$ ($M_{\odot}$) & $15.65\pm 1.45$  \\
\hline
\end{tabular} 
\caption{Selected parameters for M33 X-7.  The uncertainties
correspond to one standard deviation.  For the determination of
$\Theta$, see the Supplementary Information.  The measurements of
$T_{\rm eff}$ and $V_{\rm rot}\sin i$ were derived directly from the
spectra.  $R_2$ and $\log L_2$ were derived from the temperature,
apparent magnitude, extinction, and the distance.  The next nine
parameters were determined by fitting the radial velocity curve and
light curves simultaneously using the ELC code\cite{oro00}.  The final
two parameters are fixed by those given above.}
\end{table}

\newpage
\spacing{1}

\begin{figure}
\centerline{\includegraphics[scale=0.65,angle=-90]{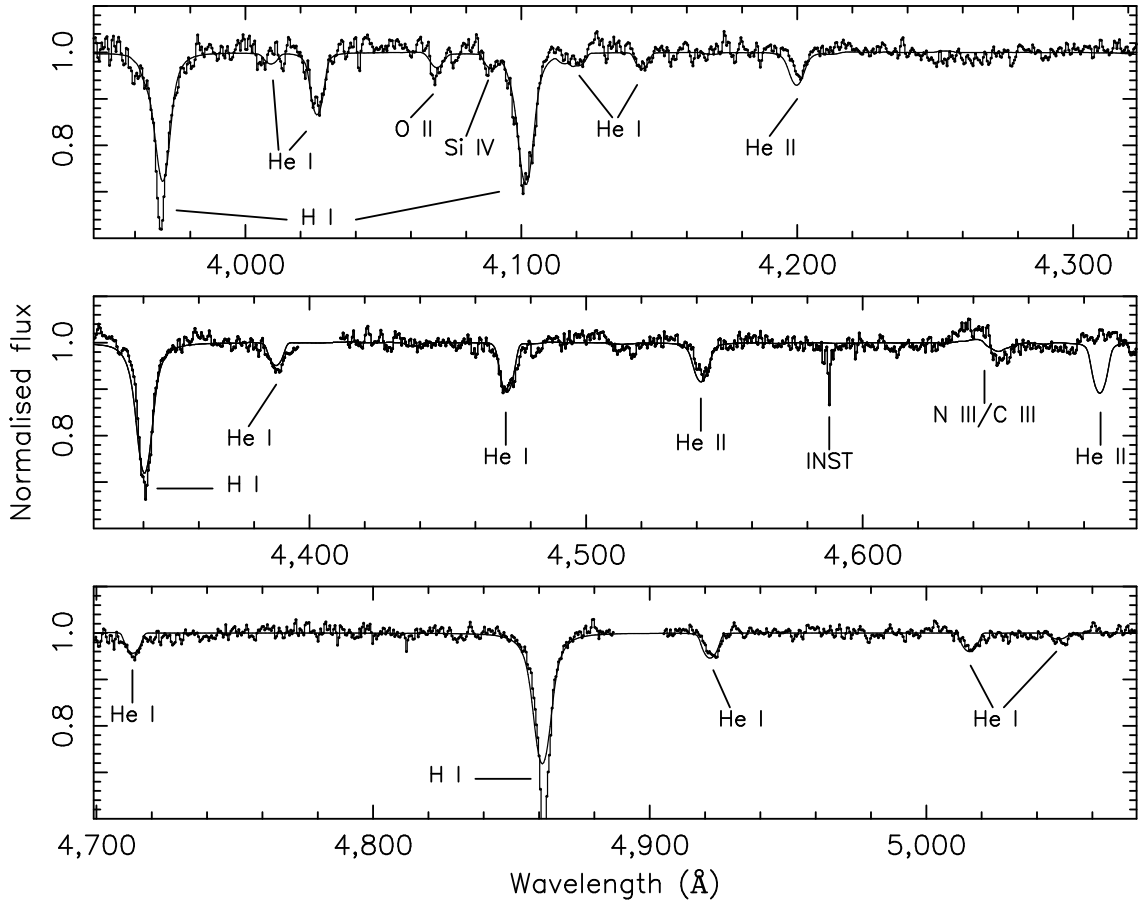}}
\caption{Mean optical spectrum of M33 X-7.  The spectrum shown here
(connected dots), which was extracted with the SPECRES package in
IRAF\cite{luc03}, is the sum of the 22 individual spectra that have
been velocity-shifted to the rest frame of the secondary star.  The
solid line is the model spectrum described in the text.  The data were
obtained using the GMOS instrument on the 8.2m Gemini-North Telescope
with the B1200 grating ($\lambda_c=4650$~\AA) and a 0.5 arcsecond slit
rotated to a position angle of 215.6 degrees, which was defined by M33
X-7 and a nearby pair of stars 0.9 arcseconds to the southwest (see
Supplementary Information).  Twenty-four 40-minute spectra were
acquired in service mode between 2006 August 18 and November 16 in
good seeing of always $< 0.8$ arcseconds.  The two observations
obtained on 2006 September 17 are suspect and will not be considered
here.  The initial bias subtraction, flat-fielding, and wavelength
calibrations were performed using the GMOS package in IRAF.  In the
two-dimensional spectra, the overlap of the profiles of M33 X-7 and
the nearby pair of stars was modest.  The optimal extraction of
one-dimensional spectra was done two ways.  (1) Routines in the GMOS
package were used with the spectral extraction aperture adjusted so
that light from the nearby pair of stars was not included, which
resulted in at most about a 20\% loss of light from M33 X-7 (see
Supplementary Information).  The final extracted spectra had
signal-to-noise ratios of 20 or more per 0.47~\AA\ pixel near
H$\beta$.  Numerous nebular emission lines from the surrounding HII
region\cite{hum80} are seen in these spectra, including the Balmer
lines H$\beta$ through H$\epsilon$, [O III] near 4363, 4969 and
5007~\AA, and weak He I lines near 4026, 4471, 4921 and 5015~\AA.  The
He II line near 4686~\AA\ and the N III lines near 4640~\AA\ are also
in emission.  The quality of the wavelength stability was checked by
measuring the radial velocity of the brightest nebular line, [O III]
5007\AA.  Its average heliocentric velocity in the 22 spectra is
$-131.2 \pm 1.5$ km\,s$^{-1}$ (std.\ dev.); for comparison, the
velocity of M33 in the NASA Extragalactic Database is $-179 \pm 3$
km\,s$^{-1}$.  (2) Routines in the SPECRES package were used to
deblend the spatial profiles of M33 X-7 and the nearby pair of stars
and to remove the nebular lines before optimally extracting
one-dimensional spectra.  However, the resulting spectra have lower
signal-to-noise ratios than the spectra extracted with the GMOS
routines (see Supplementary Information).
}
\end{figure}

\newpage

\begin{figure}
\centerline{\includegraphics[scale=0.85,angle=0]{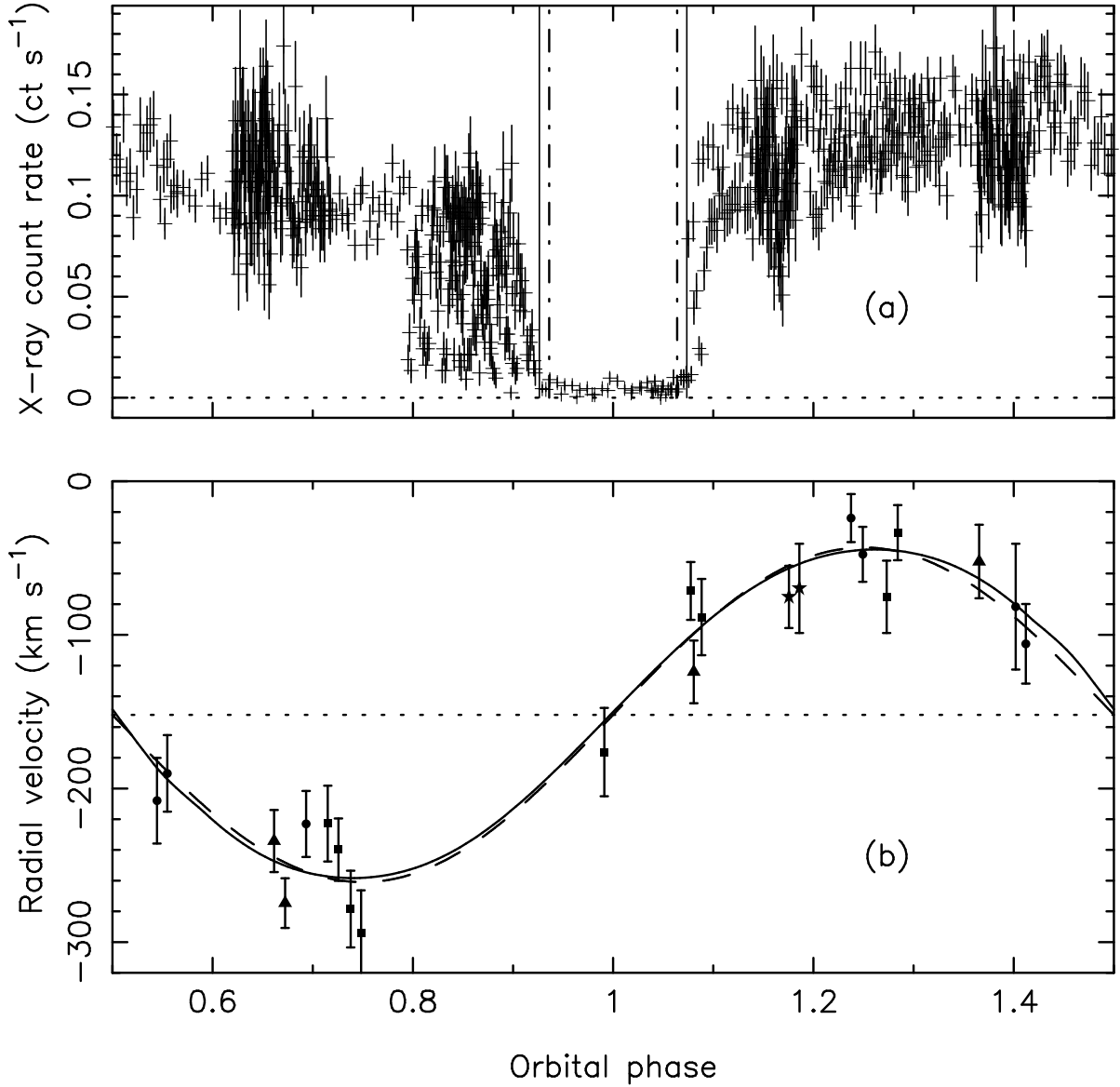}}
\caption{Phased X-ray light curve and radial velocity curve for M33
X-7.  (a) The {\em Chandra} ACIS light curve in the $0.5-5$ keV energy
band.  The error bars $1\sigma$ (s.d.) statistical.  Complete orbital
phase coverage is achieved here using all 17 available ACIS
observations, including five observations (ObsIDs 1730, 6376, 6385,
6387 and 7344) not present in Figure 1 of ref.\ 5.  The count rates
are corrected for vignetting and for the difference between the
responses of the ACIS-S and ACIS-I detectors for the single ACIS-I
observation (ObsID 6378).  The solid vertical lines denote an X-ray
eclipse duration of $\Theta=53^{\circ}$ (ref.\ 5), which incudes the
effects of an extended wind from the O-star.  The dash-dotted vertical
lines denote an eclipse duration of $\Theta=46^{\circ}$, which
corresponds solely to an eclipse by the photosphere of the O-star.
(b) The radial velocity curve derived from the Gemini spectra
(extracted using the GMOS IRAF package) with the best-fitting model
shown as a solid line.  The error bars $1\sigma$ (s.d.) statistical.
The dashed line is the best-fitting sinusoid.  The radial velocities
were derived by cross-correlating the spectra against a synthetic
spectrum (Fig.\ 1) over the wavelength ranges 4150--4300 and
4521--4578~\AA.  These bands include two He II lines, 4200\AA~and
4541\AA, which are uncontaminated by nebular lines.  Radial velocities
obtained in 2006 August are denoted by circles, 2006 September by
squares, 2006 October by stars, and 2006 November by triangles.  Using
the orbital period of 3.453014 days determined from the X-ray
eclipses\cite{pie06}, a sine fit to the 22 velocities yields $K_2 =
108.9 \pm 6.4$ km\,s$^{-1}$, systemic velocity $\gamma = -152 \pm 5$
km\,s$^{-1}$, and $T_0 = {\rm HJD}\, 2,453,967.157 \pm 0.048$.  Here,
$T_0$ is the predicted time of mid-X-ray eclipse, which is in full
agreement with that of ref.\ 5 -- they differ by $95.001 \pm 0.014$
orbital cycles.  The value of the mass function, which is the absolute
minimum mass of the compact object, is $f(M) = 0.46 \pm
0.08\,M_{\odot}$.
}
\end{figure}

\newpage

\begin{figure}
\centerline{\includegraphics[scale=0.7,angle=-90]{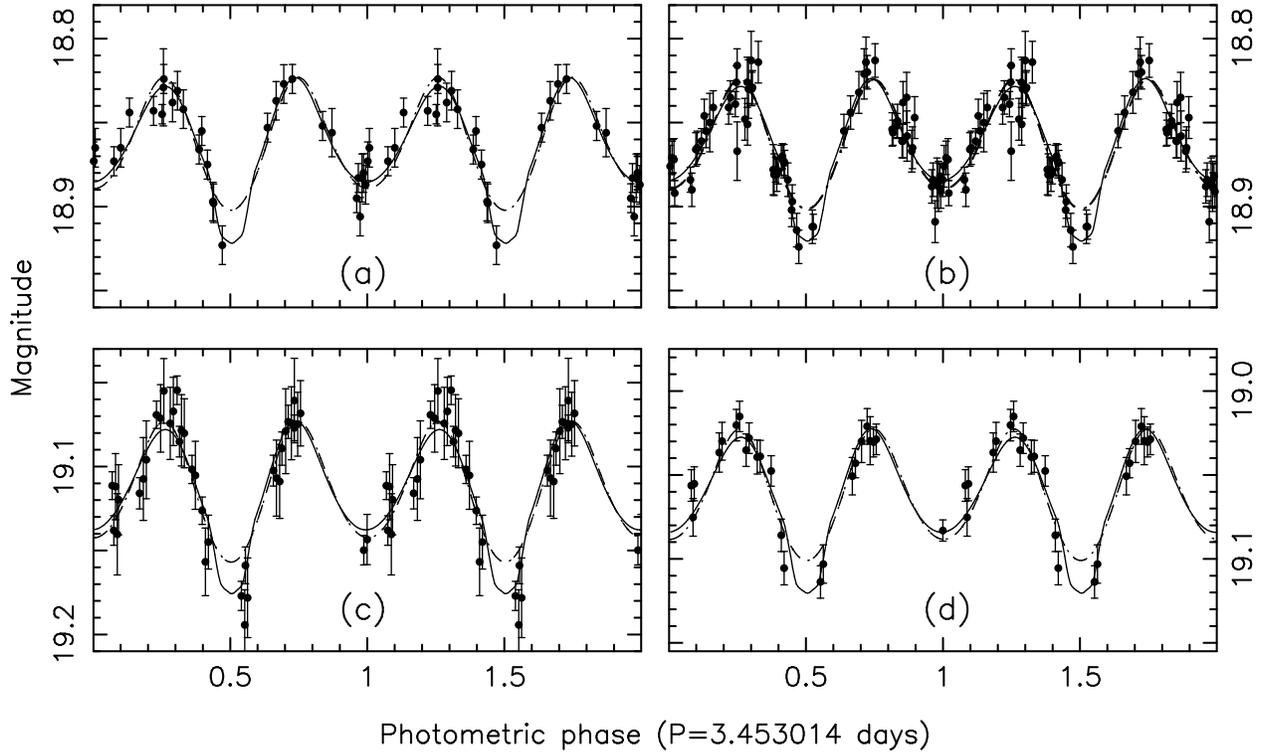}}
\caption{Optical light curves.  (a) The $B$ light curve from ref.\
10.  (b) The $V$ light curve from ref.\ 10.  (c) Gemini g$^{\prime}$
light curve.  (d) Gemini r$^{\prime}$ light curve.  The photometric
time series were derived using the ISIS image subtraction
software\cite{ala00} together with DAOPHOT\cite{ste87}, which was used
to determine the reference flux.  The error bars are $1\sigma$ (s.d.)
statistical.  
The ELC code\cite{oro00} was used to find the optimal binary model.
These light curves and
the radial velocities shown in Fig.\ 2b were used as input data.
In addition, we have three
other constraints: the radius
of the O-star of $R_2=19.6 \pm 0.9\,R_{\odot}$, the projected
rotational velocity of the O-star of $V_{\rm rot}\sin i=250 \pm 7$ km
s$^{-1}$, and the width of the X-ray eclipse of $\Theta=46\pm
1^{\circ}$ (see Supplementary Information).  We note that
the eclipse duration and
known radius are strong constraints that are unavailable for Galactic
black hole binaries.
In
deriving the models, we initially fitted for six parameters: $i$,
$K_2$, 
$M_2$ (see the Table 1 caption for definitions), $R_2$, $T_{\rm
eff}$, and a phase shift $\Delta\phi$, which is used to account for
small uncertainties in the ephemeris.  The first three of these
parameters, along with the established orbital period $P$, determine
the scale of the binary, including the dimensions of the Roche
equipotential lobes.  The value of $R_2$ then determines the
Roche-lobe filling factor $f_2$.  With the geometry of the star fully
specified, $T_{\rm eff}$ and the gravity darkening law ($T\propto
g^{1/4}$) determines the distribution of temperatures over the surface
the of the star.  No parameterised limb darkening is required because
we computed the specific intensities from the OSTAR2002 grid.
Likewise, X-ray heating has been accounted for, and is anyway a minor
correction ($\Delta T\le 100$K) because of the star's extreme
luminosity.  After several initial trial runs, we found that the fits
were improved by (1) adding a faint accretion disk around the compact
object with a fractional radius $r_d$, (2) allowing the orbit to be
slightly non-circular (adds eccentricity $e$ and argument of
periastron $\omega$ as free parameters), and (3) allowing the O-star
to rotate slightly non-synchronously with the orbit, which is
parameterised by $\Omega=P_{\rm orb}/P_{\rm rot}$ (we assume the
star's rotation axis is perpendicular to the orbital plane). 
The solid lines show the best-fitting model, and
the dash-dotted lines show the
best-fitting models with a circular orbit and no accretion disk. 
The
genetic optimiser code was run five times with different initial
random parameter sets and the grid search optimiser was run many
hundreds of times to refine the solution and define confidence limits
on the fitted and derived parameters (see Supplementary Information).
}
\end{figure}

\newpage
\begin{figure}
\centerline{\includegraphics[scale=0.75,angle=0]{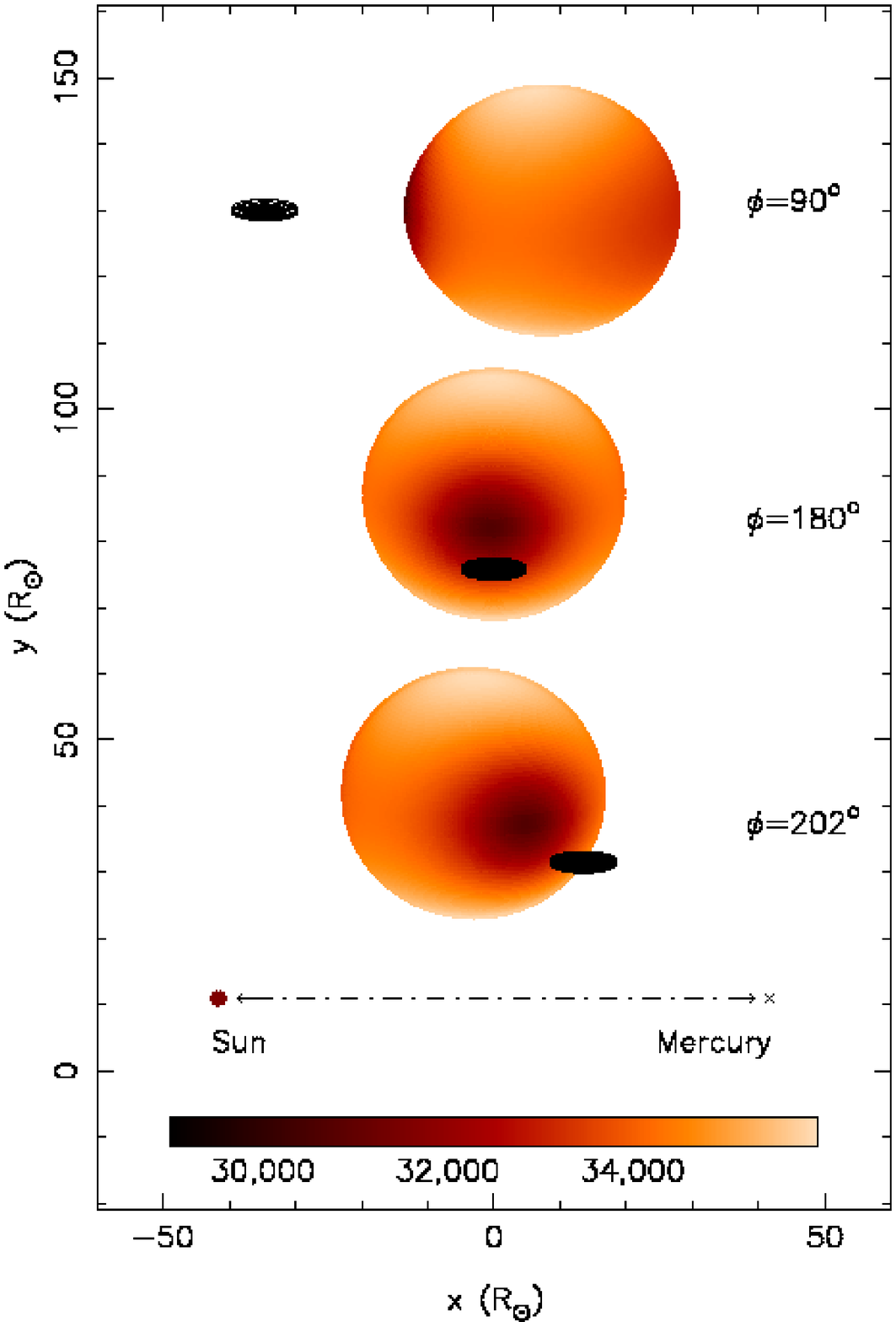}}
\caption{Schematic diagram of M33 X-7.  The companion star and the
accretion disk surrounding the black hole are shown to scale as seen
projected onto the plane of the sky at three orbital phases.  The
colours on the star represent temperatures (not intensities), with
cooler temperatures shown by darker colours as denoted on the bar.
The distance between the Sun and Mercury is indicated and the figure
is scaled in solar radii.
\label{schem}}
\end{figure}

\spacing{1}


\begin{thebibliography}{1}
\bibitem{rem06}
 Remillard, R. A. \& McClintock, J. E. X-Ray properties of black-hole binaries.
 {\em Annu.\ Rev.\  Astron.\  Astrophys.} {\bf 44}, 49--92
 (2006). 

\bibitem{cha06}
Charles, P. A. \& Coe, M. J. 
in {\it Compact Stellar X-ray
   Sources} (eds.\ W. H. G. Lewin \& M. van der Klis) 215--265
   (Cambridge University Press, Cambridge, 2006).

\bibitem{oro03}
Orosz, J. A. 
in {\em A Massive Star Odyssey:
   From Main Sequence to Supernova} (eds.\ van der Hucht, K. A., 
   Herrero, A. \& Esteban, C.) Proc.\ IAU Symp.\ 212
   365--371 (ASP, San Francisco, 2003).


\bibitem{bro01}
Brown, G. E.
Heger, A.
Langer, N.
Lee, C. -H.
Wellstein, S., \& Bethe, H. A.  Formation of high mass X-ray black 
hole binaries.
{\em New Astron.} {\bf 6}, 457--470 (2001).

\bibitem{pie06}
 Pietsch, W. {\em et al.} M33 X-7: ChASeM 33 reveals the first eclipsing 
 black hole X-Ray binary.  {\em Astrophys.\ J.} {\bf 
 646}, 420--428 (2006).


\bibitem{tau06}
Tauris, T. M. \& van den Heuvel, E. P. J. in
{\it Compact Stellar X-ray
   Sources} (eds.\ W. H. G. Lewin \& M. van der Klis)  623--665
   (Cambridge University Press, Cambridge, 2006).

\bibitem{sch92}
Schaller, G., Schaerer, D., Meynet, G., \& Maeder, A. 
New grids of stellar models from 0.8 to $120\,M_{\odot}$
at $Z=0.020$ and $Z=0.001$.
{\em Astron.\ Astrophys.\ Suppl.} {\bf 96}, 269--331 (1992).

\bibitem{mey94}
Meynet, G., Maeder, A., Schaller, G., Schaerer, D.,
\& Charbonnel, C. Grids of massive stars with high mass 
loss rates. V. From 12 to 120 $M_{\odot}$ at 
Z=0.001, 0.004, 0.008, 0.020 and 0.040.
{\em Astron.\ Astrophys.\ Suppl.} {\bf 103}, 97--105 (1994).

\bibitem{vaz07}
{{V{\'a}zquez}, G.~A.,
{Leitherer}, C.,
{Schaerer}, D., {Meynet}, G. \& {Maeder}, A.}
Models for massive stellar populations with rotation.
{\em Astrophys.\ J.} {\bf 663}, 995--1020
(2007).

\bibitem{pie04}
 Pietsch, W., Mochejska, B. J., Misanovic, Z., Haberl, F., 
 Ehle, M. \& Trinchieri, G. The eclipsing massive X-ray binary M33 X-7: 
 New X-ray observations and optical identification. {\em Astron.\ Astrophys.}
 {\bf 413}, 879--887 (2004).

\bibitem{shp07}
  Shporer, A., Hartman, J.,  Mazeh, T., \& Pietsch, W.
  Photometric analysis of the optical counterpart of the 
  black hole HMXB M33 X-7. {\em Astron.\ Astrophys.} {\bf 
  462},  1091--1095 (2007).

\bibitem{lan03}
 Lanz, T. \& Hubeny, I. A Grid of non-LTE line-blanketed model
 atmospheres of O-type stars. {\em Astrophys.\ J. Suppl.} {\bf
 146}, 417--441 (2003).


\bibitem{ma01}
Ma, J. {\em et al.} Spectral energy distributions, ages, and
   metallicities of star clusters in M 33.
   {\em Astron.\ J.} {\bf 122}, 1796--1806 (2001).

\bibitem{hea07}
Heap, S. R., Lanz, T., \& Hubeny, I. Fundamental properties of
O-type stars.  {\em Astrophys.\ J.} {\bf 638}, 409--432 (2006).


\bibitem{oro00}
 Orosz, J. A. \& Hauschildt. P. H. 
 The use of the NextGen model atmospheres for 
 cool giants in a light curve synthesis code. {\em Astron.\ Astrophys.}
 {\bf 364}, 265--281 (2000). 

\bibitem{gei03}
Geis, D. R. 
in {\em A Massive Star Odyssey:
   From Main Sequence to Supernova} (eds.\ van der Hucht, K. A., 
   Herrero, A. \& Esteban, C.) Proc.\ IAU Symp.\ 212
   91--100 (ASP, San Francisco, 2003).

\bibitem{gre01}
Greiner, J., Cuby, J. G., \& McCaughrean 2001.  An unusually massive
   stellar black hole in the galaxy.  {\it Nature} {\bf 414,} 522--525
   (2001).

\bibitem{har04}
Harlaftis, E. T. \& Greiner, J. The rotational broadening and the mass
of the donor star of GRS 1915+105. 
{\em Astron.\ Astrophys.} {\bf 414}, L13--L16 (2004).

\bibitem{nei07}
Neil, E. T., Bailyn, C. D., \& Cobb, B. E.  Infrared monitoring of
   the microquasar GRS 1915+105: detection of orbital and superhump
   signatures, {\it Astrophys.\ J.} {\bf 657,} 409--414 (2007).

\bibitem{rey97}
 Reynolds, A. P., Quaintrell, H., Still, M. D., Roche, P.,
   Chakrabarty, D. \& Levine, S. E.  A new mass estimate for Hercules
   X-1. {\it Mon. Not. R. Astron. Soc.} {\bf 288,} 43--52 (1997).

\bibitem{pod03}
Podsiadlowski, Ph., Rappaport, S., \& Han, Z.  On the formation and
evolution of black hole binaries. {\em Mon.\ Not.\ R. Astron.\ Soc.}
{\bf 341}, 385--404 (2003).


\bibitem{luc03}
 Lucy, L. B. \& Walsh, J. R. Iterative techniques 
 for the decomposition of long-slit spectra. {\em Astron. J.} {\bf 
 125},  2266--2275 (2003).

\bibitem{hum80}
 Humphreys, R. M. \& Sandage, A. On the stellar content and 
 structure of the spiral galaxy M 33.  {\em Astrophys.\ J. Suppl.}
 {\bf 44}, 319--381 (1980).

\bibitem{ala00}
 Alard, C. Image subtraction using a space-varying kernel.
 {\em Astron.\ Astrophys.\ Suppl.} {\bf 144},  363--370 (2000). 

\bibitem{ste87}
 Stetson, P. B. DAOPHOT - A computer program for crowded-field 
stellar photometry.  {\em Pub.\ Astron.\ Soc.\ Pac.} {\bf 99}, 191--222
(1987).

\end{thebibliography}
\end{document}